\documentclass[fleqn,usenatbib]{mnras}

\usepackage[english]{babel}
\usepackage{newtxtext,newtxmath}
\usepackage{adjustbox}
\usepackage{multirow}
\usepackage{makecell}
\usepackage{subfig}
\usepackage{xcolor}
\usepackage{caption}
\usepackage{hyperref}
\usepackage{multicol}

\usepackage[T1]{fontenc}

\DeclareRobustCommand{\VAN}[3]{#2}
\let\VANthebibliography\thebibliography
\def\thebibliography{\DeclareRobustCommand{\VAN}[3]{##3}\VANthebibliography}

\usepackage{graphicx}	
\usepackage{amsmath}	

\title[Redshift Difference Cosmology]{The Redshift Difference in Gravitational Lensed Systems:\\ A Novel Probe of Cosmology}

\author[C. Wang et al.]{
Chengyi Wang,$^{1}$
Krzysztof Bolejko$^{2}$ and
Geraint F. Lewis$^{1}$
\\
$^{1}$Sydney Institute for Astronomy, School of Physics, A28, The University of Sydney, NSW 2006, Australia\\
$^{2}$School of Natural Sciences, College of Sciences and Engineering, University of Tasmania, Private Bag 37, Hobart, TAS 7001, Australia
}

\date{Accepted XXX. Received YYY; in original form ZZZ}

\pubyear{2023}

\begin{document}
\label{firstpage}
\pagerange{\pageref{firstpage}--\pageref{lastpage}}
\maketitle

\begin{abstract}
The exploration of the redshift drift, a direct measurement of cosmological expansion, is expected to take several decades of observation with stable, sensitive instruments.
We introduced a new method to probe cosmology which bypasses the long-period observation by observing the \emph{redshift difference}, an accumulation of the redshift drift, in multiple-image gravitational lens systems.
With this, the photons observed in each image will have traversed through different paths between the source and the observer, and so 
the lensed images will show different redshifts when observed at the same instance. 
Here, we consider the impact of the underlying cosmology on the observed redshift difference 
in gravitational lens systems, generating synthetic data for realistic lens models and exploring the accuracy of determined cosmological parameters.
We show that, whilst the redshift difference is sensitive to the densities of matter and 
dark energy within a universe, it is independent of the Hubble constant.
 Finally, we determine the observational considerations for using the redshift difference as a cosmological probe, finding that one thousand lensed sources are enough to make robust determinations of the underlying cosmological parameters. {Upcoming cluster lens surveys, such as the Euclid, are expected to detect a sufficient number of such systems.}
\end{abstract}


\section{INTRODUCTION}
As a famous prediction of general relativity, cosmological expansion has been extensively explored over in the past century. The first observable clue, published by \cite{1913LowOB...2...56Slipher}, found the redshift of light from a distant galaxy that interpreted these galaxies as receding away from the Earth. Later, the theory of expansion was established by \cite{1922ZPhy...10..377Friedmann} and \cite{1927ASSB...47...49Lematre}, which was finally confirmed by \cite{1929PNAS...15..168Hubble}. During the past half-century, various kinds of cosmological observations have achieved remarkable results, including CMB \citep{2003ApJS..148...97Bennett,2007ApJS..170..288Hinshaw_Bennett}, type Ia supernovae \citep{2004ApJ...607..665Riess}, and the large-scale distribution of galaxies \citep{2001Natur.410..169Peacock, 2005ApJ...633..560Eisenstein}, which have provided accurate results and convenience, {opening an era of } "precision cosmology" \citep{2005NewAR..49...25Primack}. One of the most unexpected discoveries in the last three decades was that the Universe continues accelerating its expansion \citep{1998AJ....116.1009Riess}. The physical law behind this phenomenon, until the present, is still uncertain. One of the most accepted hypotheses of this dark energy is Einstein's cosmological constant $\Lambda$ \citep{1992ARA&A..30..499Carroll}, a constant component with $w = -1$ in the equation of state. Of course, $w$ can actually be time-dependent \citep{1997MNRAS.289L...5Chiba,2000IJMPD...9..373Sahni}. Generally, the varying $w$ component may lead to an inhomogeneous universe with $-1 < w < 0$ \citep{krasinski_1997}. On the contrary, another alternative effort to explain the acceleration is to add dark matter and modify gravity \citep{2002PhRvD..65d4023Deffayet}. However, most of the observations agree with the results of the cosmological constant. The most convenient and feasible way to determine the physical reasons that dictate acceleration is a directly examine the expansion history \citep{2007MNRAS.382.1623Balbi}. In practice, classical methods for retracing the expansion history involve observing type Ia supernovae \citep{2004ApJ...607..665Riess} and baryon acoustic oscillations \citep{2007ApJ...665...14Seo}. \par

Unlike other cosmological observables that rely on the distance, such as SN Ia, redshift drift \citep{1961ApJ...133..355S_Sandage} is a direct measurement of the expansion history. Because the galaxies in our universe are not stationary but drift along with the Hubble flow, the redshift of galaxies changes with the universal expansion. We have followed the redshift drift papers by \cite{1961ApJ...133..355S_Sandage,1962ApJ...136..319S_Sandage}, as well as other relevant works discussed by \cite{Loeb_1998,2008MNRAS.386.1192Liske,2008PhRvD..77b1301Uzan,2012PhR...521...95Quercellini, Cooke_2019,2020EPJC...80..304Liu} who have provided a complete summary of the possibility of measuring the redshift drift in a decade observational time interval with the next-generation telescope. \par

In addition to the challenge of measuring redshift drift, observing this phenomenon requires several decades between two epochs. To address this issue, we have developed a new method for measuring the \emph{redshift difference} in gravitational lenses. Our focus has been on gravitational lens systems that can produce multiple images of a celestial object, as shown in Figure 1 of our previous paper \citep[hereafter Paper I]{2022ApJ...940...16Wang}. Two photons are emitted from a distant source at different times and converge at the telescope simultaneously. If the time delay between the two images is longer than two decades, the redshift difference between them will be the same as the redshift drifts over ten years. This redshift difference can be detected using next-generation telescopes like the Extreme Large Telescope (ELT). However, not all lens systems are suitable for detecting the redshift difference. We have found that choosing a cluster rather than a galaxy as a lens can provide a sufficient time delay between the images. Furthermore, we expect better results for detecting the redshift difference with sources with larger redshift values.\par

We have also tested several different dark energy models for the redshift difference, including (1) the simplest $\Lambda \rm{CDM}$ model, which we have used in our work and explained in our results; (2) a slight extension of $\Lambda \rm{CDM}$ called the $w \rm{CDM}$ model \citep{1999ApJ...517..565Perlmutter}, in which dark matter is not a constant component over time; and (3) the Chevalliear-Polarski-Linder (CPL) model \citep{PhysRevLett.90.091301Linder}, which is another extension of the $\Lambda \rm{CDM}$ model that includes $w(a) = w_0 + w_a (1-a)$, giving the equation of state two degrees of freedom. Using these three cosmological models, we have tested the redshift difference in non-singular isothermal ellipsoidal (NIE) lens systems to constrain the cosmological parameters.\par

We have structured this paper as follows. In Section \ref{sec: redshift difference}, we provide background information on redshift drift and redshift difference. In Section \ref{sec: cosmological constraints}, we introduce the main procedures for analyzing the redshift difference. We also present the constraints on cosmological parameters derived from the redshift difference in Section \ref{sec: constraints on cosmological mdoels}. Finally, we summarize our work in Section \ref{sec: summary}.

\section{THE REDSHIFT DIFFERENCE}
\label{sec: redshift difference}

\subsection{The redshift difference in a stationary lens system}
Due to the gravitational lensing effect, the trajectory of light from a distant source is curved, resulting in a slightly different redshift compared to the source without a lens in between. As a result, the redshift measured in the lensed images should be different, giving rise to the "redshift difference" \citep{2022ApJ...940...16Wang}. This method offers a significant advantage over the redshift drift as it does not require a waiting period of 20 years to observe the difference.\par

This redshift difference arises due to a direct result of the gravitational lens. Assuming the lens is at an angular diameter distance of $D_l$, the source at $D_s$ and the angular diameter distance in between is $D_{ls}$. The deflection angles of light are represented by the Greek letters $\alpha$ (for the lens), $\beta$ (for the source), and $\theta$ (for the image). {The photons in two different images} are emitted from a distant source at two different instances and received at the telescope instantaneously. For more information, you can refer to Paper I. The lens equation is as follows \citep{Schneider_1, Schneider_2}:
\begin{align}
    \begin{split}
        \boldsymbol{\beta} = \boldsymbol{\theta} - \frac{D_{ls}}{D_{s}}\boldsymbol{\alpha},
    \end{split}
    \label{eq: lens equation}
\end{align}
By solving Eqn. \ref{eq: lens equation}, we can determine the positions of the images, given the position of the source. For example, if we already have two image position $\boldsymbol{\theta_1}$ and $\boldsymbol{\theta_2}$, according to \cite{1986ApJ...310..568Blandford}, the time delay at source between them is:
\begin{align}
    \begin{split}
        \Delta t_{12} = \frac{D_l D_s}{D_{ls}}\frac{1+z_l}{1+z_s} \left[\tau({\boldsymbol{\theta_1}},{\boldsymbol{\beta}})-\tau({\boldsymbol{\theta_2}},\boldsymbol{{\beta}}) \right],
    \end{split}
\end{align}
where $\boldsymbol{\theta_i}$ is the image position of $i$-th image. Because the redshift difference is regarded as a tiny correction of the redshift in each image, which means we have $z(\boldsymbol{\theta_1}) - z(\boldsymbol{\theta_2}) \approx dz(\boldsymbol{\theta_1})/dt_0 \times \Delta t_{12} \approx dz_s/dt_0 \times \Delta t_{12}$. Accurate to the first order of the redshift difference, we obtain
\begin{align}
    \begin{split}
        \Delta z_{12}  \approx   -(1+z_l) H(z_s) \frac{D_l D_s}{c D_{ls}} \left[ \tau({\boldsymbol{\theta_1}};\boldsymbol{\beta})-\tau(\boldsymbol{\theta_2};\boldsymbol{\beta}) \right].
    \end{split}
    \label{eq: redshift difference}
\end{align}
{One of the most important property we can infer from Eqn. \ref{eq: redshift difference} is that the first order of redshift difference is independent of the Hubble constant $H_0$. This is because the Hubble function $H(z_s)$ is proportional to $H_0$, while the angular diameter distances $D_s$, $D_l$, and $D_{ls}$ are proportional to $H_0^{-1}$.}\par

\subsection{The Discussion of the peculiar velocity}

Before further argument of the redshift difference, we should also consider the peculiar motion of the lens system, which can impact the redshifts in the images. {Therefore, it is reasonable to split the redshift of $i$-th image into several parts
\begin{align}
    \begin{split}
        1 + z_{i} \equiv (1 + z_{s}) \times (1 + z_{cosmo}) \times (1 + z_p)
    \end{split}
\end{align}
where $z_{s}$ refers to the redshift if there is no gravitational lens, $z_{cosmo} \ll 1$ is the slight redshift due to the gravitational lens, and the $z_{p} \ll 1$ is the redshift due to peculiar velocity of the lens system. So the redshift difference between two images could be written as, by subtracting the $z_i$ and $z_j$, $\Delta z_{ij} / (1 + z_s) \approx \Delta z_{cosmo} + \Delta z_p$, where $\Delta z_{cosmo}$ is the redshift difference between image $i$ and image $j$ if the lens system is static and $\Delta z_p$ is the redshift difference due to peculiar velocity.} Further more, the redshift difference due to peculiar velocity can be expanded into signals arising from peculiar motion of observer, lens and source. In this subsection, we will discuss that the redshift differences due to the source's and observer's velocities are negligible, whereas the effect from the lens's peculiar velocity is of a similar order of magnitude as the cosmological redshift and can be determined.\par

Firstly, the redshift rising from the source peculiar velocity can be composed as $z_{source} = v_{source} / c$ \citep{2014MNRAS.442.1117Davis}, where $c$ is the speed of light. As a result, the redshift difference due to the source's peculiar velocity, to the first order of redshift difference, is 0. Only when peculiar acceleration is under consideration is the redshift difference due to the source non-zero. According to \cite{2008PhLB..660...81Amendola}, the appearance of peculiar acceleration is around $\sim 1 \rm{cm/s}$ per decade which equals to a drift of redshift $\sim 10^{-11}$ per year. It is at least 2 orders smaller than the redshift difference between the lensed images we discussed above. Based on the similar reason, the result of the observer's peculiar velocity is negligible. {\cite{2017GrCo...23..240Feoli_local_acceleration} estimated the local acceleration is around $1.4\times 10^{-9} \rm{m/s^2}$, that is equivalent a redshift difference of $\sim 10^{-10}$ if the time interval of two images' measurements is as long as half year. Thus we can ignore the local acceleration since its effect is two to three order smaller than the signal.}\par

Unfortunately, in most circumstances, we cannot overlook the lens' velocity, because it will directly affect the lens equations when bending the trajectory of photons. However, if we can detect the redshift at the level of $1\ \rm{cm/s}$, it is possible to determine the redshift difference due to the lens's peculiar velocity. Accepting the conclusions from \cite{1983Natur.302..315Birkinshaw}, the change of redshift is composed as
\begin{align}
    \begin{split}
        z_{lp} = B\ \gamma\ \sin\delta\ \cos\phi\ \alpha
    \end{split}
\end{align} 
where $B = v_{lens}/c$, $\gamma = \sqrt{1 - B^2}$ and $\delta$ is the angle between the lens velocity and the line of sight. Solving the lens equations for each source, the lens equation reads $\beta = \theta_i - \frac{D_{ls}}{D_s} \alpha_i$, where the subscript $i$ refers to the $i$-th image, the deflection angle $\alpha$ is a function of $\beta$, $\theta$ and lens parameters, $\alpha = \alpha(\theta; \beta, \pi_{\small lens})$ assuming the $\pi_{\small lens}$ is the lens configuration. Therefore the redshift due to lens peculiar velocity transforms into $z_{lp} = B \gamma \sin(\delta) \cos(\phi) \alpha(\theta; \beta, \pi_{\small lens})$. Considering the image position, source position, and lens parameters, we can figure out the lens' velocity in the image plane. Normally, the redshift due to the lens' peculiar velocity is around $10^{-7} \sim 10^{-6}$ \citep{2021JCAP...09..018Dam, 2010MNRAS.402..650Kill}. 

\begin{figure}
    \centering
    \includegraphics[width=\columnwidth]{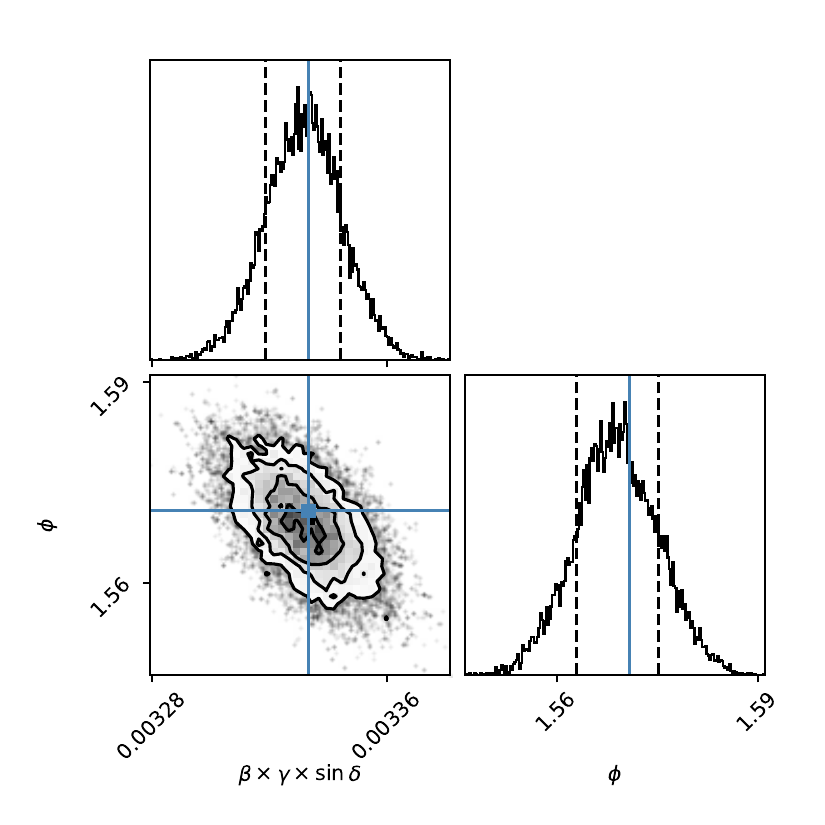}
    \caption{A sample of lens velocity in the image plane of $10$ sources in a cluster lens with velocity dispersion $\sigma_v = 1500 \rm{km/s}$ where the angle $\phi$ is the direction of lens velocity in the image plane, the $\delta$ is the angle between lens velocity and line of sight, $B = v_{lens}/c$ and $\gamma=\sqrt{1-B^2}$.}
    \label{fig:n10}
\end{figure}

If there is  only one source in the lens system, the error is about $10^{-8}$, only redshift difference in very rich clusters with velocity dispersion $\sigma_v \sim 1500\ \rm{km/s}$ is capable of be determined. However, if more than one sources exist in a typical lens system, we can constrain the peculiar redshift more precisely. We discussed in \cite{2022ApJ...940...16Wang} that the error of redshifts due to peculiar velocity can be restricted to $10\%$ if there are more than $10$ sources. To demonstrate this, we generated mock observables taking the accuracy of the High-Resolution Spectrograph (HIRES) from \cite{2021Msngr.182...27M_HIRES}, in which we assume the error of position is $\sim 3\ \rm{mas}$ and the accuracy of redshift is $\sim 6\ \rm{cm/s}$. Then we used \texttt{emcee} \citep{Foreman_Mackey_2013emcee} and \texttt{lenstronomy} \citep{2018PDU....22..189Blenstronomy} to sample the lens' peculiar velocity in the image plane. As an example, we show the result of $10$ sources in a cluster lens system in Fig. \ref{fig:n10}. With $10$ sources in a lens system, we constrain the accuracy to $10^{-9}$ for the cosmological redshift difference $10^{-8} \sim 10^{-7}$. Thus we only consider the cosmological redshift difference in next Sec. \ref{sec: cosmological constraints}.

\section{SUMMARY OF ANALYSIS}
\label{sec: cosmological constraints}
In this section, we provided a summary of the analysis, including lens modelling and analysis to determine cosmological parameters from lens models assuming each cluster owning single source in the lens system. We adopted the goals outlined HIRES of ELT from \cite{2021Msngr.182...27M_HIRES}, which require a position resolution of around $1\ \rm{mas}$ and a spectrum resolution $\sim 6\ \rm{cm/s/decade}$ that meets the requirements of Sandage's test \citep{2014JCAP...12..018Geng}. \par

{To illustrate the detectability of redshift difference by metal absorption lines in V band using the ELT, we considered an extreme scenario in which the source is at redshift $z_s = 4$ with the lens mass $M \approx 3 \times 10^{15} M_\odot$ whose redshift difference reaches $\Delta z \sim 10^{-7}$.  We can regard the spectral difference of two images as a tiny intensity change. \cite{2001A&A...374..733Bouchy} established a method of improving the radial velocity measurements. In each given $i$-th pixel, the spectrum shift is defined as
\begin{align}
    \begin{split}
        A_2(i) - A_1(i) = \frac{\partial A(i)}{\partial \lambda(i)} \delta \lambda(i) 
        = \frac{\partial A(i)}{\partial \lambda(i)} \frac{\delta v(i)}{c} \lambda(i)
    \end{split}
\end{align}
where $\delta v(i)/c = \delta \lambda/\lambda$. So the velocity shift can be expressed as
\begin{align}
    \begin{split}
        \frac{\delta v(i)}{c} = \frac{A_2(i) - A_1(i)}{\lambda(i)\frac{\partial A(i)}{\partial \lambda(i)}}
    \end{split}
\end{align}
With the estimation of pixel noise, one can determined the velocity shift accuracy
\begin{align}
    \begin{split}
        \sigma_v(i) = \frac{c \sigma(i)}{\lambda(i)\frac{\partial A(i)}{\partial \lambda(i)}}
    \end{split}
\end{align}
where $\sigma(i) = [A_2(i) - A_1(i)]_{\rm{RMS}}$. Therefore, the average velocity uncertainty is \citep{2001A&A...374..733Bouchy, 2022AJ....164...84Artigau}
\begin{align}
    \begin{split}
        \sigma_v = \frac{1}{\sqrt{\sum \sigma_v(i)^{-2}}}.
    \end{split}
    \label{eq: velocity uncertainty}
\end{align}
We utilized the Python package \texttt{bagpipes} \citep{2018MNRAS.480.4379C_bagpipes} to generate galaxy spectral lines in the V band at three different spectrum resolutions: R = 200,000, 100,000, and 50,000. Fig. \ref{fig: sigmaSNRT} illustrates the variation of velocity uncertainty with signal-to-noise ratio (S/N) for each spectral resolution, as obtained from Equation \ref{eq: velocity uncertainty} with a source-magnitude $\sim 20$. In order to detect a redshift difference of $10^{-7}$, a velocity uncertainty of 100 cm/s is required. Panels (a), (b), and (c) depict the reduction in velocity uncertainty as the S/N increases for each resolution, respectively. The dash blue reference lines represent the critical condition for detecting the signal, while the green (red) areas indicate the permissible (forbidden) regions; Panel (d) shows the total observational time for each spectral resolution. The \texttt{ANDES Exposure Time Calculator} \footnote{\url{http://hires.inaf.it/etc.html}} was used for this purpose with the same telescope efficiency as \cite{2008MNRAS.386.1192Liske}. The red, orange, and yellow lines correspond to resolutions of $200,000$, $100,000$, and $50,000$ respectively. The dashed lines represent the critical conditions, which are the same as the blue dashed lines in panels (a), (b), and (c). Additionally, the blue dashed-point line represents the requirement for measuring the redshift drift as outlined by \cite{2008MNRAS.386.1192Liske}. With a $1000$-hour observation, the high S/N allows us to collect enough photons to detect a large redshift difference.\par

\begin{figure*}
    \flushleft
    \hspace*{-0.0cm}
    \subfloat[\label{fig:a}]{
		\includegraphics[scale=0.55]{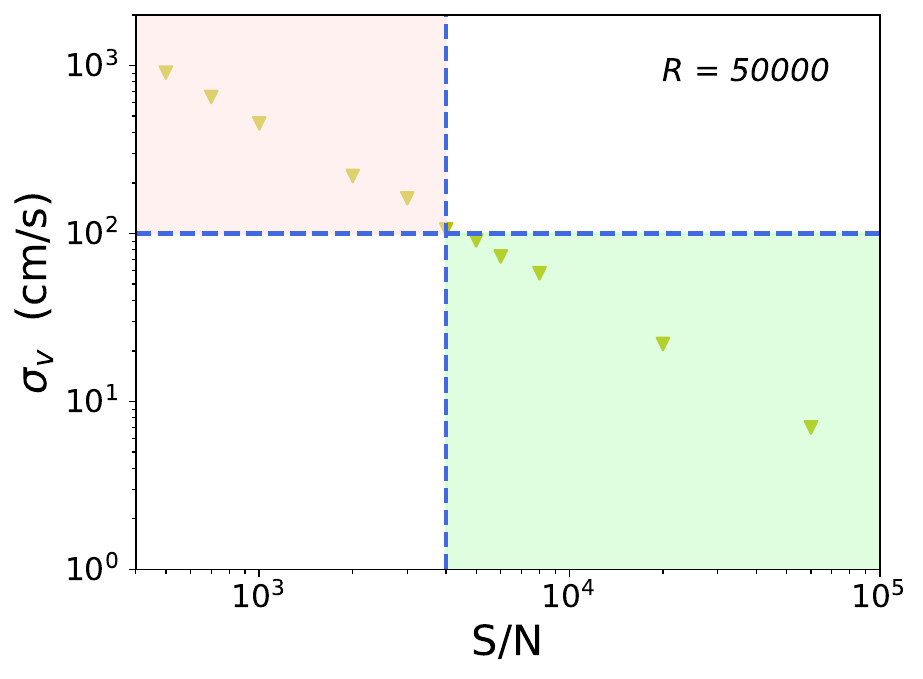}}
    \subfloat[\label{fig:b}]{
		\includegraphics[scale=0.55]{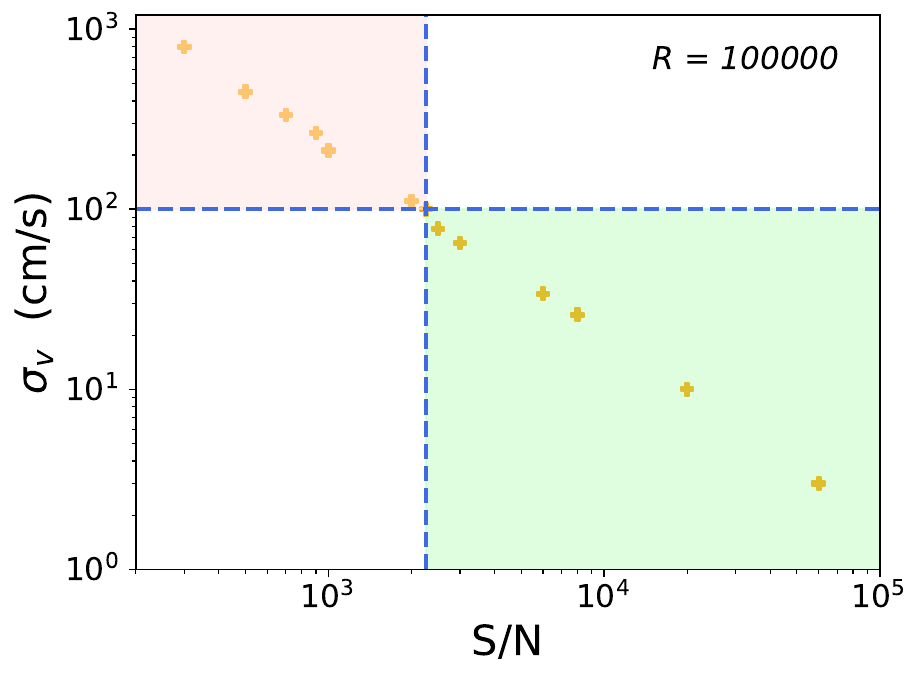}}
    \\
    \hspace*{-0.0cm}
	\subfloat[\label{fig:c}]{
        \includegraphics[scale=0.55]{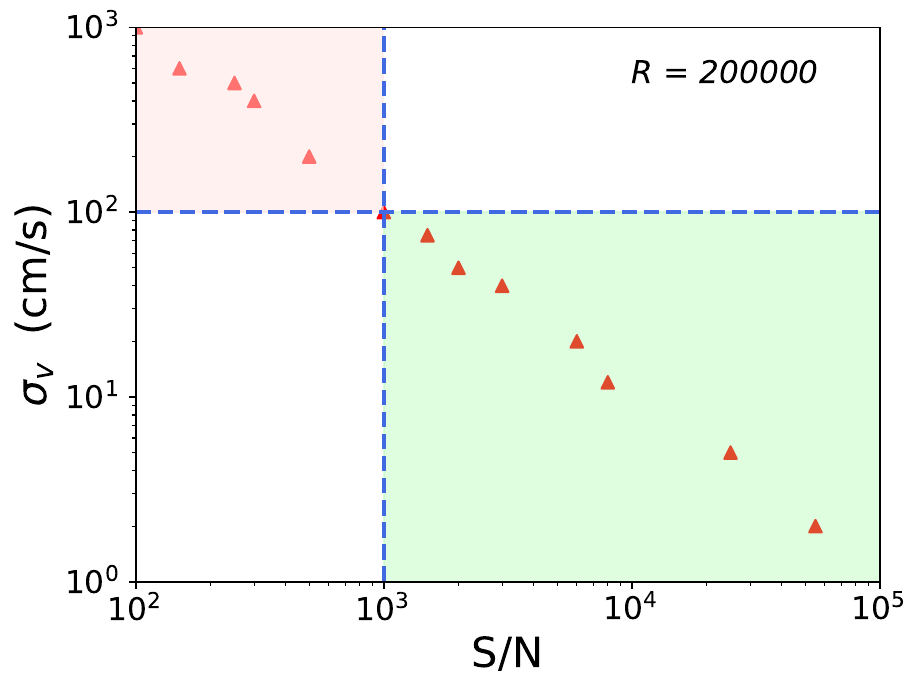}}
    \subfloat[\label{fig:d}]{
        \includegraphics[scale=0.55]{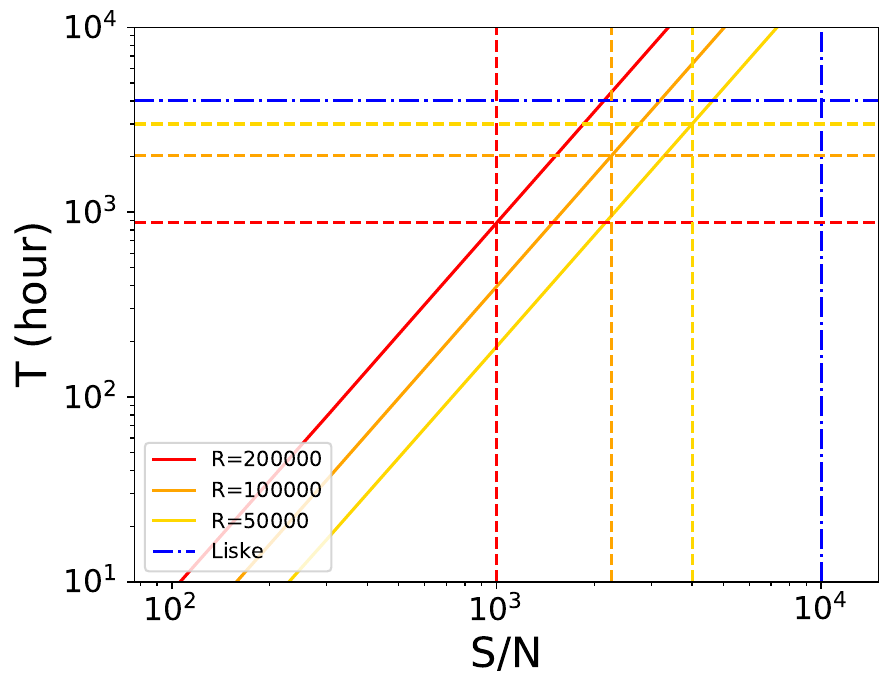}}
    
    \caption{The variation of velocity uncertainty with signal-to-noise ratio (S/N) for each spectral resolution, as obtained from Equation \ref{eq: velocity uncertainty} with a source-magnitude $\sim 20$. In order to detect a redshift difference of $10^{-7}$, a velocity uncertainty of 100 cm/s is required. Panels (a), (b), and (c) depict the reduction in velocity uncertainty as the S/N increases for each resolution, respectively. The dash blue reference lines represent the critical condition for detecting the signal, while the green (red) areas indicate the permissible (forbidden) regions; Panel (d) shows the total observational time for each spectral resolution. The \texttt{ANDES Exposure Time Calculator} was used for this purpose with the same telescope efficiency as Liske. The red, orange, and yellow lines correspond to resolutions of $200,000$, $100,000$, and $50,000$ respectively. The dashed lines represent the critical conditions, which are the same as the blue dashed lines in panels (a), (b), and (c). Additionally, the blue dashed point line represents the requirement for measuring the redshift drift as outlined by Liske. This indicates that a $1000$-hour observation is sufficient to detect a significant redshift difference.}
    \label{fig: sigmaSNRT}
\end{figure*}

If the sources are high redshift quasars \citep{2023arXiv230514317Napier}, we can adopted the results from \cite{2008MNRAS.386.1192Liske} by the 40-meter diameter ELT with instrument efficiency $\varepsilon \sim 0.25$. In this case, the velocity resolution is able to reach 1 cm/s. The required S/N for targets with $z < 4$ is determined by the Eqn. 16 in  \cite{2008MNRAS.386.1192Liske}:
\begin{align}
	\begin{split}
		\sigma_v = 1.35\ \left( \frac{S/N}{2370}\right)^{-1}\left( \frac{N_{QSO}}{30} \right)^{-1/2}\left( \frac{1+z}{5} \right)^{-1.7}\ \rm{cm/s}
	\end{split}
\end{align}
where $S/N$ is the signal-noise ratio, $N_{QSO}$ is the absorption lines, and $z$ is the redshift of source. For sources $z>4$, the exponent over redshift is $-0.9$. The spectral $S/N$ is dominant by four parameters:
\begin{align}
	\begin{split}
		S/N = 700 \left[ \frac{Z_X}{Z_r} 10^{-0.4(16-m_X)}
			\left(\frac{D}{42\ \rm{m}} \right)^2\ 
			\left( \frac{t_{ini}}{10\ \rm{h}} \right)\ 
			\left( \frac{\epsilon}{0.25} \right)  \right]^{1/2}
	\end{split}
\end{align}
where $Z_X$ and $m_X$ are the zero-point and magnitude of the source, the $Z_r$ is the zero-point of AB band, the $D$, $t_{ini}$, $\epsilon$ are diameter of the telescope, integral time and efficiency of the instruments. We provide a quick estimate of the required $S/N$ for high-redshift quasars, using the results from previous studies \citep{2003AJ....125.1649Fan, 2012MNRAS.420.1764Roche}. For high-redshift quasars with a redshift of about $z \simeq 6$ and an apparent magnitude of $z_{AB} \simeq 20$, the $S/N$ is expected to be around $\sim 1600$, assuming a telescope diameter of $D=40\ \rm{m}$, an initial exposure time of $t_{ini}=2000\ \rm{h}$, and a spectrograph efficiency of $\epsilon=0.25$. Therefore, achieving the required $S/N$ is feasible with future observations. 

Notably, the redshift difference method enables the instantaneous measurement of redshift in two images, thereby circumventing the instability issues often associated with spectrographs. This approach differs from the conventional method, which entails measuring redshift drift over a 20-year time interval. The ANDES aims for wavelength calibration stability, targeting a precision of $2\ \rm{cm/s}$ over a 10-year period \citep{2022arXiv221202458Milakovic, 2022SPIE12184E..24Marconi}. Consequently, the instrument's stability would not pose a concern during the half-year observation required for the redshift difference measurements.
}

\subsection{Lens modelling}
As there are no existing redshift difference measurements, we have used ten artificial lenses to demonstrate the lens models and for further redshift difference analysis. We have chosen the NIE model \citep{1998ApJ...495..157Keeton, 1994A&A...284..285Kormann, 2019A&A...622A.165Delchambre, 1995ApJ...441...58Wallington_Christopher} to build our lens models, whose 2-D surface density is:
\begin{align}
    \begin{split}
        \tau(x, y) = \frac{\theta_E}{2} \left({s^2 + q x'^2 + \frac{y'^2}{q}} \right)^{-1/2}
    \end{split}
    \label{eq: surface density}
\end{align}
In this equation, $\theta_E$ represents the Einstein radius, and $s$ corresponds to the size of the lens core. However, not all gravitational lenses are suitable for measuring redshift differences. Typically, galaxies are too small to generate a vast enough redshift difference to detect. For instance, a typical galaxy lens with a velocity dispersion of $\sigma_v \sim 100 \rm{km/s}$ leads to a redshift difference of around $10^{-10}$. A rich galaxy cluster with a velocity dispersion of $\sigma_v \sim 1500 \rm{km/s}$ leads to a redshift difference of approximately $10^{-7}$. Therefore, it is more likely to detect the redshift difference by measuring the cluster lensing, and we will assume that the lenses are galaxy clusters in the following discussion.\par

Finally, we should confirm the impact of the lens configuration on the redshift difference. We set the error of the lensing parameters to be around $10\%$ \citep{2008A&A...477..397Grillo,2020MNRAS.498.1420H0LiCOWXIII}. The uncertainty of lens parameters leads to a residual to the redshift difference of approximately $10^{-8}$. Hence, we add an Gaussian error of $10^{-8}$ to the redshift difference obtained from the lens model.\par

\begin{figure}
    \centering
    \includegraphics[width=0.5\textwidth]{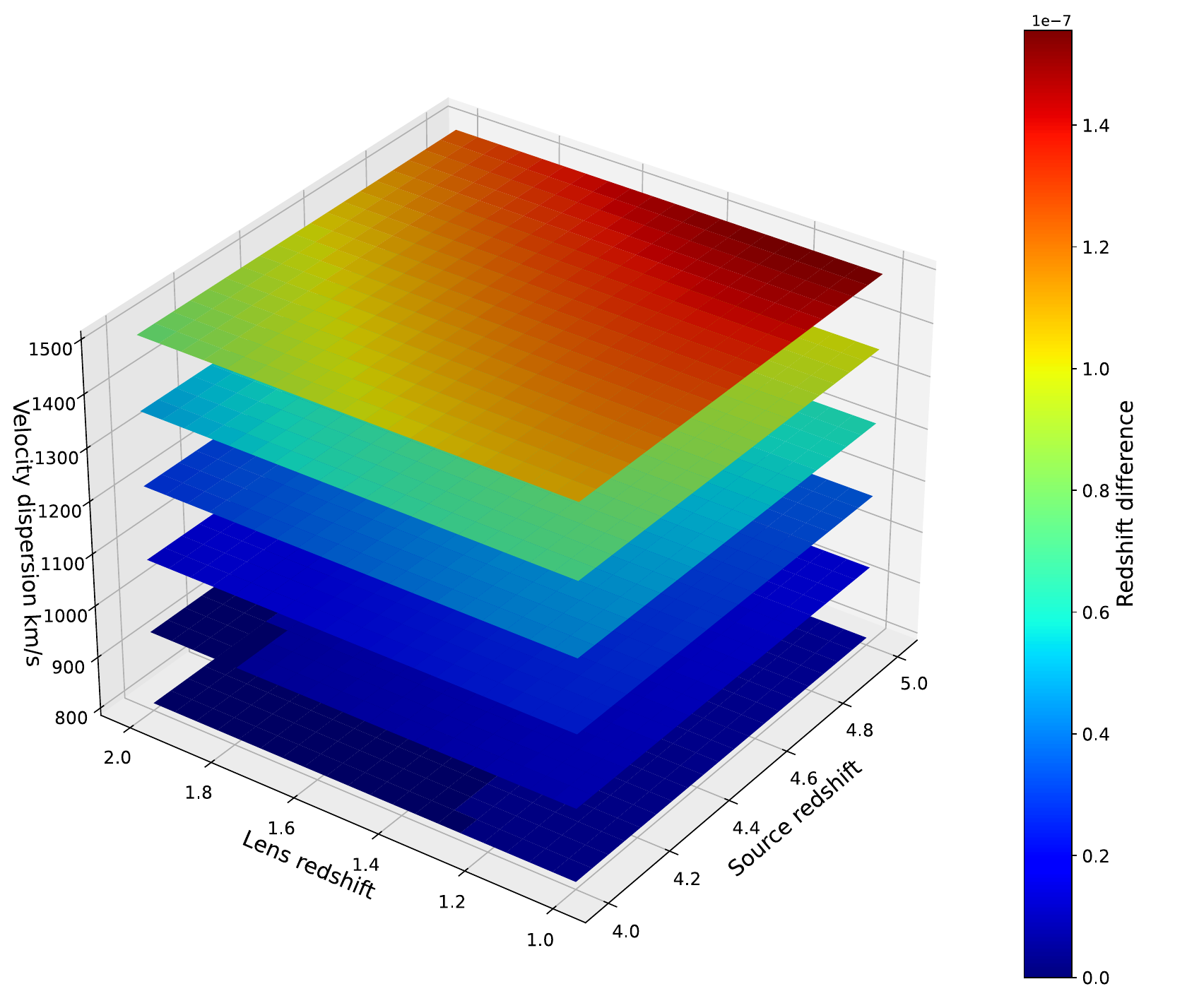}
    \caption{The redshift difference variation as a function of the velocity dispersion, lens redshift, and source redshift under the NIE lens model. It is evident that the redshift difference is positively correlated with the velocity dispersion and redshift of the source. On the other hand, it is negatively correlated with the redshift of the lens.}
    \label{fig: redshift difference 3D}
\end{figure}

As shown in Fig. \ref{fig: redshift difference 3D}, we illustrate how the redshift difference varies with the velocity dispersion, lens redshift, and source redshift under the NIE lens model. It is prominent that the redshift difference has a positive correlation with the velocity dispersion and source redshift. On the other hand, it is inverse-correlated with the redshift of the lens.

\subsection{MCMC}
\label{sec:mcmc}
To obtain constraints on cosmological parameters using lens redshift difference, we used the \texttt{emcee} \citep{Foreman_Mackey_2013emcee} Python package, based on the MCMC algorithm, to perform the analysis. Specifically, we sampled the cosmological parameters in the following manner: 
\begin{itemize}
    \item[(1)] simulating non-singular isothermal elliptical lens systems using the \texttt{lenstronomy} package \citep{2018PDU....22..189Blenstronomy}. In this step, we obtained basic information about the lenses, including their redshift and lens configurations. In the NIE lens model, the primary lens parameters are $(z_l, z_s, \sigma_v, s, e_1, e_2, x', y')$, where $z_l$ and $z_s$ are the redshifts of the lens and source respectively, $\sigma_v$ is the velocity dispersion, $s$ is the core size of the lens, the letters $e$ refer to the ellipticity components, and $(x', y')$ denotes the profile centre. We select the redshift of lenses within $(0.3,0.8)$ and the source uniformly within $(3,6)$. Note that we require rich clusters, and thus we set the velocity dispersion between $1000\ \rm{km/s}$ and $2000\ \rm{km/s}$, which corresponds to the mass of lenses of the order of $10^{15}$ to $ 10^{16}\ M_\odot$ \citep{2013MNRAS.430.2638Munari}. 
    {In this section, we aim to estimate the number of potential massive cluster lenses ($M \ge 10^{15}\ M_\odot$) within the redshift range of $0.3<z<0.8$. To accomplish this, we utilize the Press-Schechter mass function \citep{2019PhRvD..99l4036Bolejko, 1974ApJ...187..425Press}:
\begin{align}
    \begin{split}
        n(M) = f(\sigma_M)\frac{\bar{\rho}_M}{M}\frac{\partial \ln \sigma_M^{-1}}{\partial M}
    \end{split}
\end{align}
where $\sigma_M$ is the variance of matter density field smoothing at $M$, $\bar{\rho}_M$ is the mean density and $M$ is the mass. The variance of density could be written as:
\begin{align}
    \begin{split}
        \sigma_M^2 = \frac{1}{2\pi^2}\int_0^\infty
         dk\ k^2P(k)W(kR)
    \end{split}
\end{align}
where $P(k)$ is the matter power spectrum and $W(kR)$ is the window function. In general, the $f(\sigma_M)$ is a multiplicity function, but we can adopt an approaching form of:
\begin{align}
    \begin{split}
        f(\sigma_M) = A e^{-c/\sigma_M^2} 
        \left[\left( \frac{b}{\sigma_M}\right)^a+B \right]
    \end{split}
\end{align}
where $A$, $B$, $a$, $b$, $c$ are functions of redshift. Therefore, the number of potential massive clusters is:
\begin{align}
    \begin{split}
        N = \int_{z_1}^{z_2} dz\frac{dV}{dz}\int_{M_{min}}^{M_{max}} dM \frac{dn(M)}{dM}
    \end{split}
\end{align}
We utilize the mass function provided by python package \texttt{hmf} to estimate the number of ultra-massive clusters within the redshift range of $0.3<z<0.8$. According to our calculations, there are approximately $257 - 728$ such clusters in this redshift range. The \texttt{hmf} package provides the $dn/dm$ functions for each mass and redshift bin, $m+dm$ and $z+dz$. By summing $dn/dm$ over each bin. Considering the minimum, middle, and maximum values of $dn/dm$ in each bin, we are going to have different number of clusters. Besides, different halo mass functions will also affect the number of clusters. We used \texttt{hmf} and checked various mass functions, including the mass function provided by \cite{2008ApJ...688..709Tinker} and \cite{1974ApJ...187..425Press}, and they provide the range of possible outcomes. Our results indicate spread of numbers between 257 and 728. Additionally, the results from \cite{2007ApJ...654..714Hennawi} suggest that while only a part of the low-mass cluster lenses ($M \le 3\times 10^{14} M_{\odot}$) shows strong lensing effects, but all of the massive clusters ($M \ge 10^{15}\ M_\odot$) act as effective strong lenses. Therefore, if each massive cluster lens contains more than one source, we can expect to have a number of RDMs in the future.}
    
    \item[(2)] Once the lens modelled, we calculated the redshift difference between the images. To simplify the analysis, we selected the maximum redshift difference among the images for discussion below; 
    
    \item[(3)] We then built up the cosmological models, $P(z_l, z_s; \pi)$, where $\pi$ is the list of cosmological parameters, and sampled the parameters using \texttt{emcee} with the likelihood function $\mathcal{L} = -\frac{1}{2} \sum (\Delta z_{mod}(z_l,z_s;\pi) - \Delta z_{obs}(z_l,z_s;\pi))^2 / \sigma^2$, where $\Delta z_{mod}$ is the redshift difference calculated from the models and $\Delta z_{obs}$ is the observed redshift difference. Here, $z_l$ and $z_s$ refer to the redshift of the lens and source, respectively, and $\pi$ denotes the cosmological parameters. The denominator contains the error of the redshift difference, $\sigma$. Throughout the analysis, we used the prior function shown in Table \ref{tab: prior function}, where $0<\Omega_M<1$, $0<\Omega_\Lambda<1$, $0<w_0<1$, and $-1<w_a<1$. We present the results of sampling under different cosmologies in the next section.    
\end{itemize}

\begin{table}
    \centering
\begin{tabular}{ |c|p{1.5cm}<{\centering}|p{1cm}<{\centering}|p{1cm}<{\centering}|p{1.6cm}<{\centering}| } 
\hline 
\  & {Flat$\Lambda$CDM} & {$ \Lambda$CMD} & {$w$CDM} & {$w_0w_a$CDM} \\ 
\hline
$\Omega_M$ & \multicolumn{4}{|c|}{$0<\Omega_M<1$} \\ 
\hline
$\Omega_\Lambda$ & $\backslash$ & \multicolumn{3}{|c|}{$0<\Omega_\Lambda<1$} \\ 
\hline
$w_0$ & $\backslash$ & $\backslash$  & \multicolumn{2}{|c|}{$-2<w_0<0$} \\ 
\hline
$w_a$ & $\backslash$ & $\backslash$  & $\backslash$  & $-1<w_a<1$ \\ 
\hline
\end{tabular}
    \caption{The prior function for the three cosmological models comprises a shared function for $\Omega_M$. The allowable range for $\Omega_M$ is $0<\Omega_M<1$ across all four models. Furthermore, the parameter for the equation of state includes the following ranges: $0<\Omega_\Lambda<1$, $0<w_0<1$, and $-1<w_a<1$.}
    \label{tab: prior function}
\end{table}

\section{CONSTRAINTS ON COSMOLOGICAL MODELS}
\label{sec: constraints on cosmological mdoels}
Before presenting the results, we must consider several cosmological models. \texttt{Astropy} \citep{2013A&A...558A..33AstropyI,2018AJ....156..123AstropyII} provides several pre-built cosmological models \citep{1999astro.ph..5116Hogg, PhysRevLett.90.091301Linder, 2001IJMPD..10..213Chevallier}, which we will use in our analysis. We will begin with the simplest $\Lambda$CDM model, followed by the $w$CDM model with $w_0 = -1$ in the equation of state. Finally, we will investigate the redshift difference in the Chevallier-Polarski-Linder model (CPL). For each part, we will present the results of the sampling by \texttt{emcee}.\par

In this section, we use the formula proposed by \cite{2021Msngr.182...27M_HIRES}, assuming an image position resolution of $1\ \rm{mas}$ and a spectrum resolution that meets the Sandage test \citep{1961ApJ...133..355S_Sandage,1962ApJ...136..319S_Sandage} for the next-generation telescope. 

\subsection{The flat {$\Lambda$CDM} model}
\label{sec: flat LCDM}
\begin{figure}
    \flushleft
    \hspace*{-0.8cm}
    \subfloat[\label{fig:a}]{
		\includegraphics[scale=0.28]{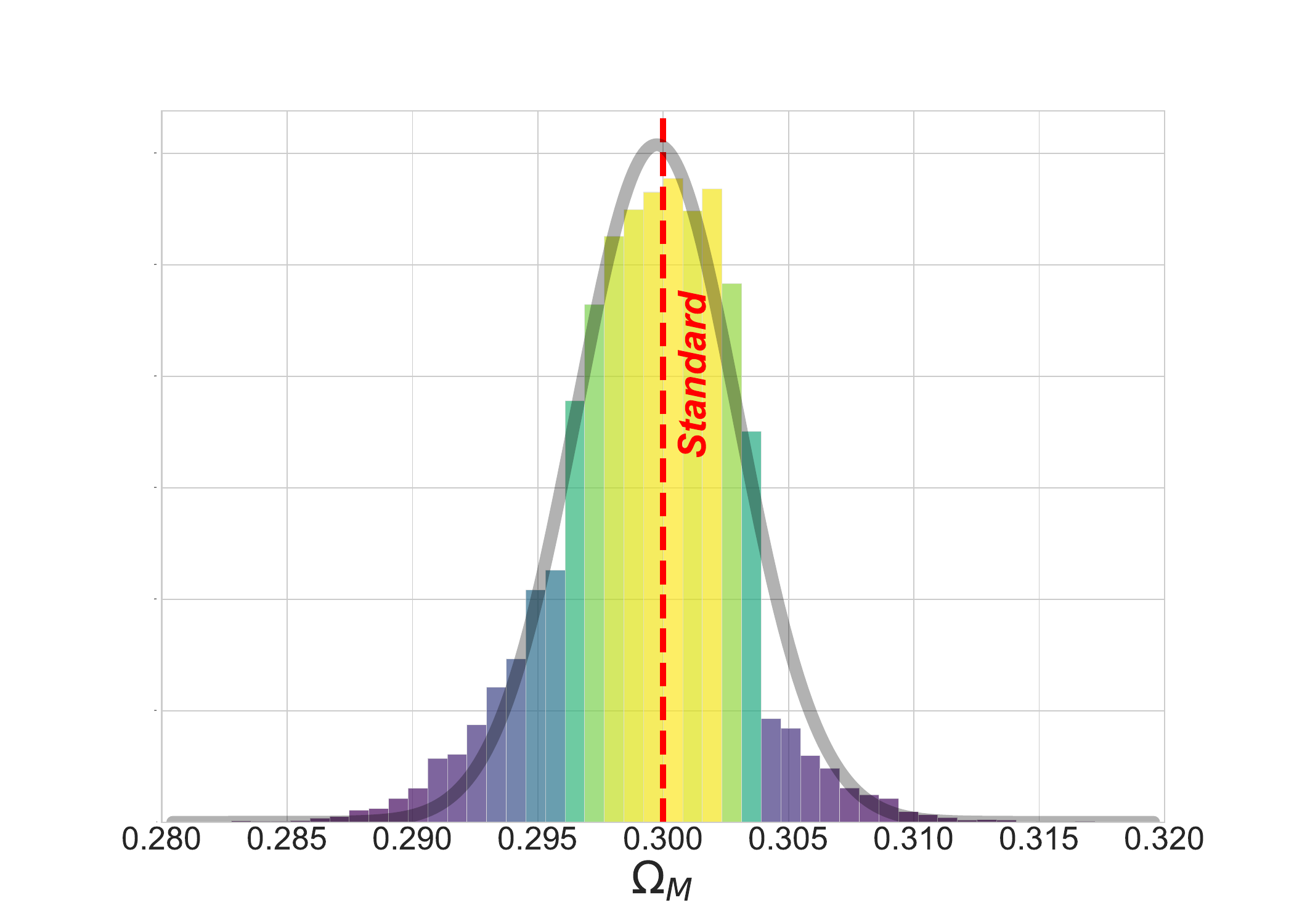}}
    \\
    \hspace*{-0.8cm}
	\subfloat[\label{fig:b}]{
        \includegraphics[scale=0.34]{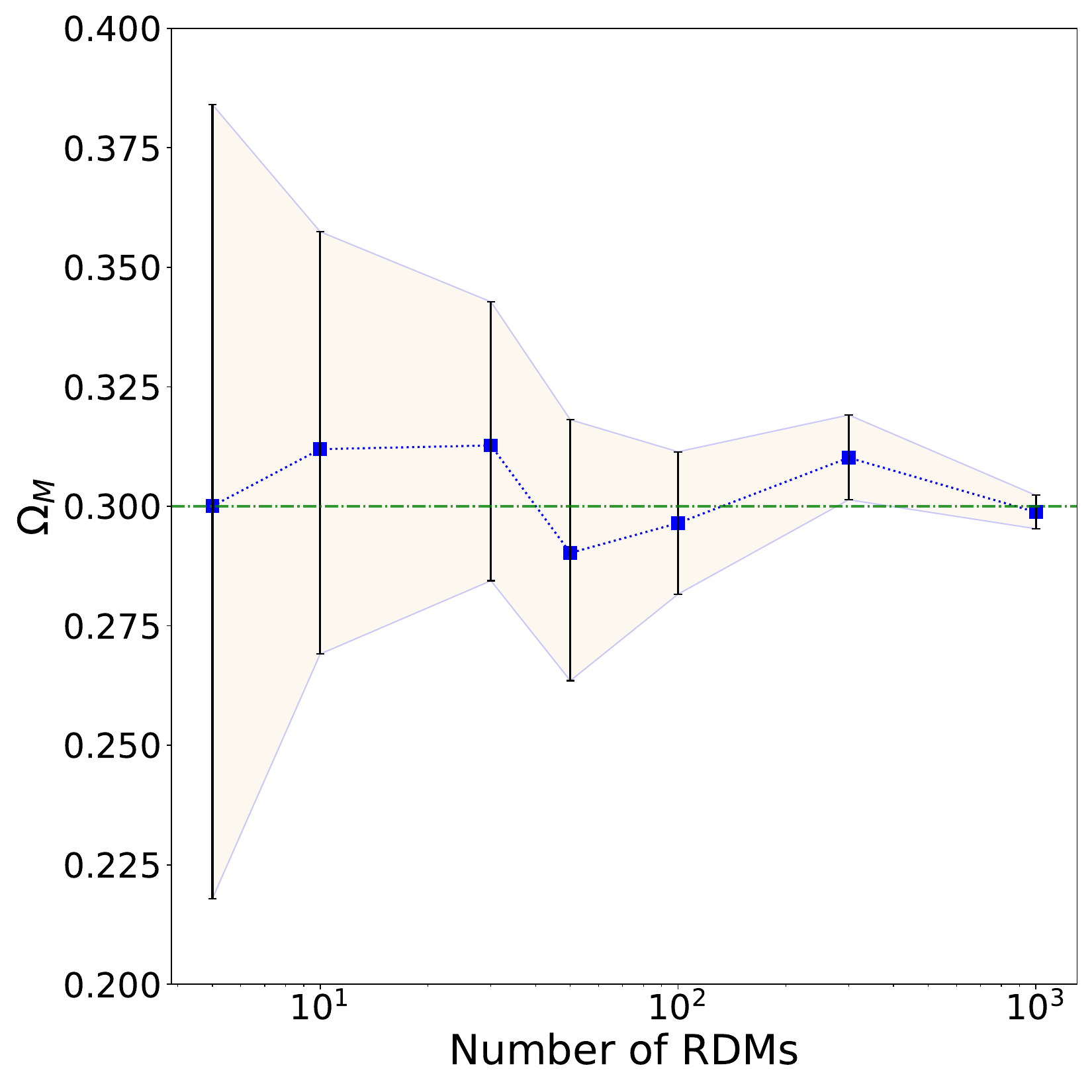} }
    
    \caption{In the upper panel of the corner plot, we show the MCMC samples obtained for the flat $\Lambda$CDM model. It is evident that the redshift difference strongly constrains the matter component $\Omega_M$. In the lower two panels, we provide an illustration of how $\Omega_M$ and $H_0$ change as more cluster lenses are included in the analysis. The red dashed line represents the true values (i.e., those of the standard model), while the black bar represents the $68\%$ confidence interval.}
    \label{fig: FlatCDM cluster number}
\end{figure}

The flat $\Lambda$CDM model is the simplest cosmological model, which assumes the existence of dark energy, dark matter, and ordinary matter. It provides satisfactory explanations for a wide range of observations, including the cosmic microwave background (CMB) \citep{1996ApJ...464L...1Bennett}, large-scale structure \citep{Lyke_2020SDSS}, Big Bang nucleosynthesis \citep{1996bboe.book.....SBBN}, and Type Ia supernovae (SN Ia) \citep{1998AJ....116.1009Riess}. We adopt the standard cosmological model where the exponent in the equation of state is a constant, $w(z) = P_\Lambda / \rho_\Lambda = -1$, with $P_\Lambda$ and $\rho_\Lambda$ denoting the pressure and density of dark energy, respectively. In addition to dark energy, we consider matter components composed of baryons ($\Omega_b$) and cold dark matter ($\Omega_{CDM}$), both of which are non-relativistic. We can write the Hubble function in a flat $\Lambda$CDM universe as:
\begin{align}
    \begin{split}
        H(z) = H_0 \left[ \Omega_M(1+z)^3 + (1-\Omega_M) \right]^{1/2}
    \end{split}
\end{align}
The results of the MCMC analysis are presented in panel (a) of Fig. \ref{fig: FlatCDM cluster number}. It is evident that the redshift difference strongly constrains the matter component $\Omega_M$ but allows for a wide range of Hubble constants. Panel (b) shows how the error of $\Omega_M$ changes as the number of lenses increases. The panel suggests that the deviation of the redshift difference shrinks rapidly when the number of cluster lenses is more than 100. To achieve more precise results, it is desirable to collect more than 500 cluster lenses.\par

\subsection{The $\Lambda \rm{CDM}$ model}
\label{sec: LCDM model}

\begin{figure}
    \centering
    \includegraphics[width=0.5\textwidth]{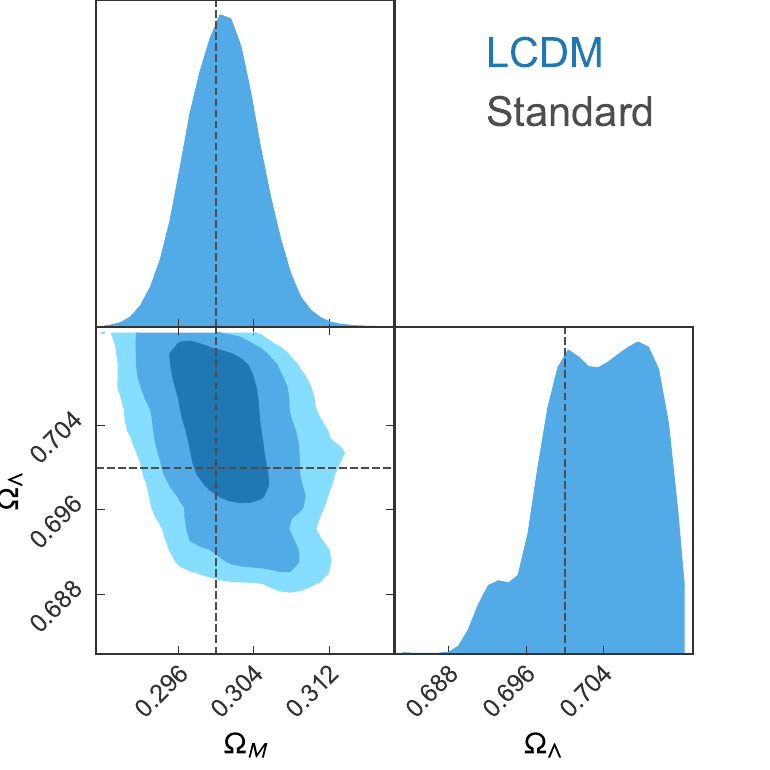}
    \caption{The MCMC result of $\Lambda$CDM for 1000 lensed sources, with the priors in Tab \ref{tab: prior function}, where the dashed line is the standard cosmology with $\Omega_M=0.3$, $\Omega_\Lambda=0.7$. $68\%$, $95\%$ and $99\%$ confidence limits {are shown as the dark, medium, and light shades of blue, respectively. }}
    \label{fig: LCDM}
\end{figure}

If we do not assume that the universe must be flat, then $\Omega_\Lambda$ becomes an independent variable in addition to $\Omega_M$, and the Hubble function takes on a different form:
\begin{align}
    \begin{split}
        H(z) = H_0 \left[ \Omega_M(1+z)^3 + \Omega_k (1+z)^2 + \Omega_\Lambda \right]^{1/2}
    \end{split}
\end{align}
The MCMC results for $\Lambda$CDM using 1000 clusters with a redshift error of $10^{-8}$ are presented in Fig. \ref{fig: LCDM}, where the dashed line represents the standard cosmology with $\Omega_M=0.3$ and $\Omega_\Lambda=0.7$. In the top panel of Fig. \ref{fig: LCDM}, the standard model lies within the $68\%$ confidence level in the $\Omega_M$ and $\Omega_\Lambda$ plane. The redshift difference estimates $\Omega_M=0.301\pm{+0.004}$ and dark energy component $\Omega_\Lambda=0.702^{+0.007}_{-0.008}$ within $68\%$ confidence level. These results are in agreement with recent observations and are even more precise since the redshift difference requires a next-generation telescope to detect.

\begin{figure}
    \centering
    \includegraphics[width=0.5\textwidth]{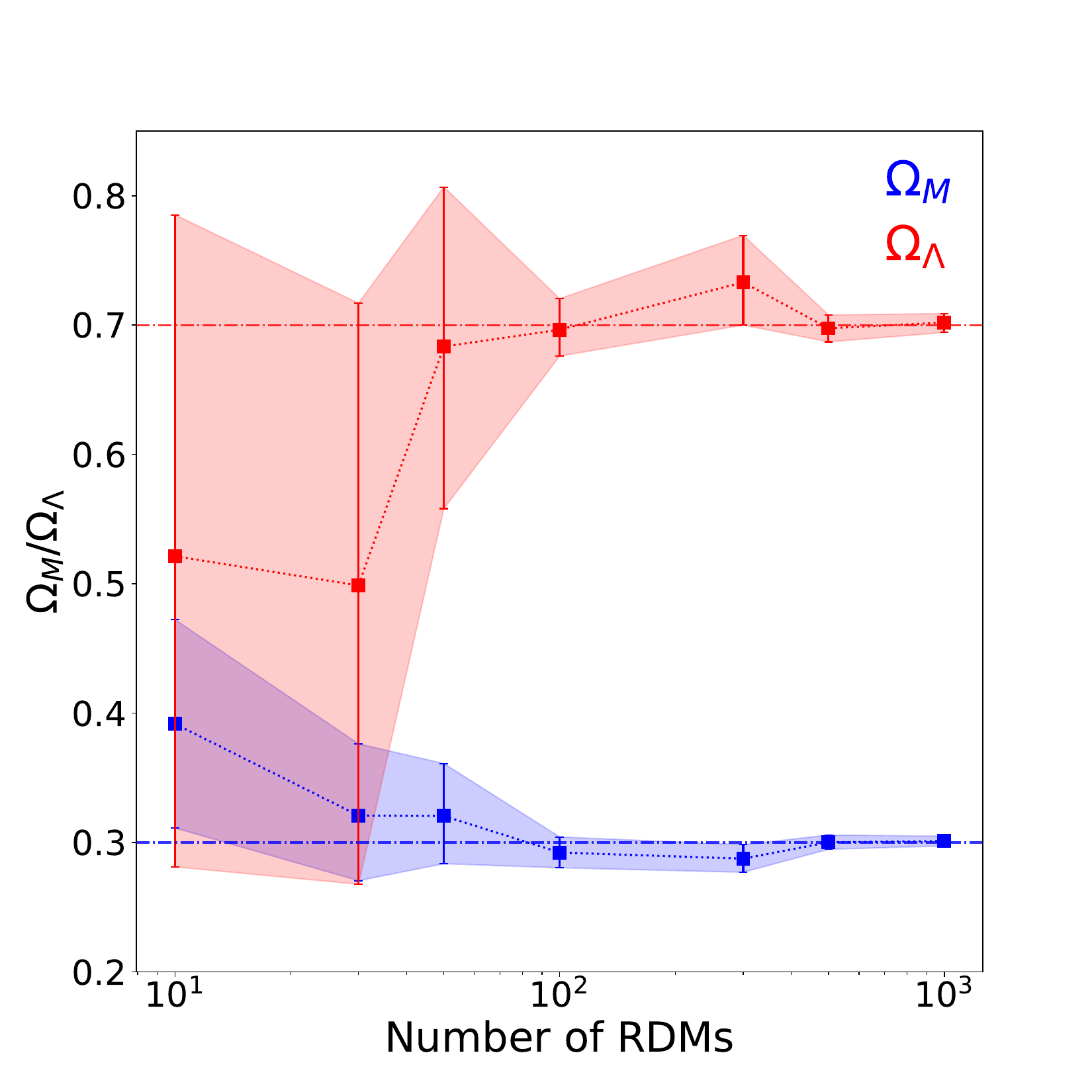}

    \caption{The square markers and dash line show the average value of the cosmological parameters with a different number of clusters, and the error bars represent the range within $68\%$ (the deep colour area in Fig. \ref{fig: LCDM}). The red dash-and-dot line represents the true value of the $\Lambda$CDM universe, where the red colour refers to the $\Omega_\Lambda$, and the blue colour refers to the matter component $\Omega_M$.}
    \label{fig: LCDM cluster number}
\end{figure}

Then, we would like to discuss the number of galaxy clusters needed in measurements. In Fig. \ref{fig: LCDM cluster number}, the blue square markers and dash line show the average value of the cosmological parameters under different numbers of measured clusters, and the black error bar is the range within $68\%$ (the shaded area in Fig. \ref{fig: LCDM}). The red dash-and-dot line represents the truthful values of $\Omega_M=0.3$ and $\Omega_\Lambda=0.7$. At small numbers of clusters, such as $10$ or $30$ clusters, the standard $\Omega_M$ is outside the $68\%$ range of the sampling data. When the number of targets is more than $500$ clusters, we obtain a narrower range of $\Omega_M$ and $\Omega_\lambda$. \par

\subsection{The flat $w$CDM model}
\label{sec: Flat wCDM}

\begin{figure}
    \centering
    \includegraphics[width=0.5\textwidth]{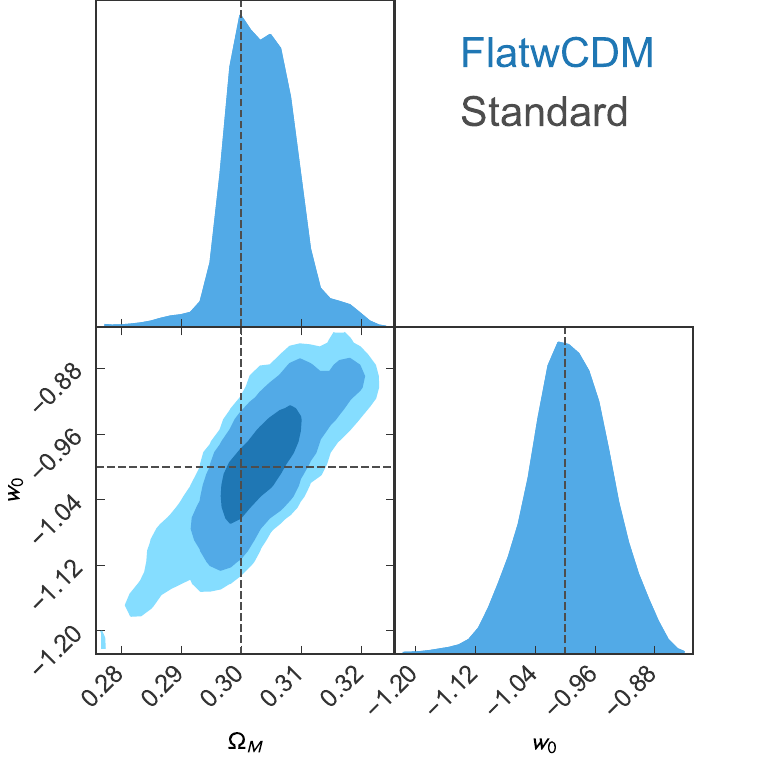}
    \caption{The contours for 1000 clusters in a flat $w$CDM model are shown at the $68\%$, $95\%$, and $99\%$ confidence levels. The black dashed lines correspond to the true values of $\Omega_M = 0.3$ and $w_0 = -1.0$.}
    \label{fig: flat wCDM}
\end{figure}

\begin{figure}
    \centering
    \includegraphics[width=0.5\textwidth]{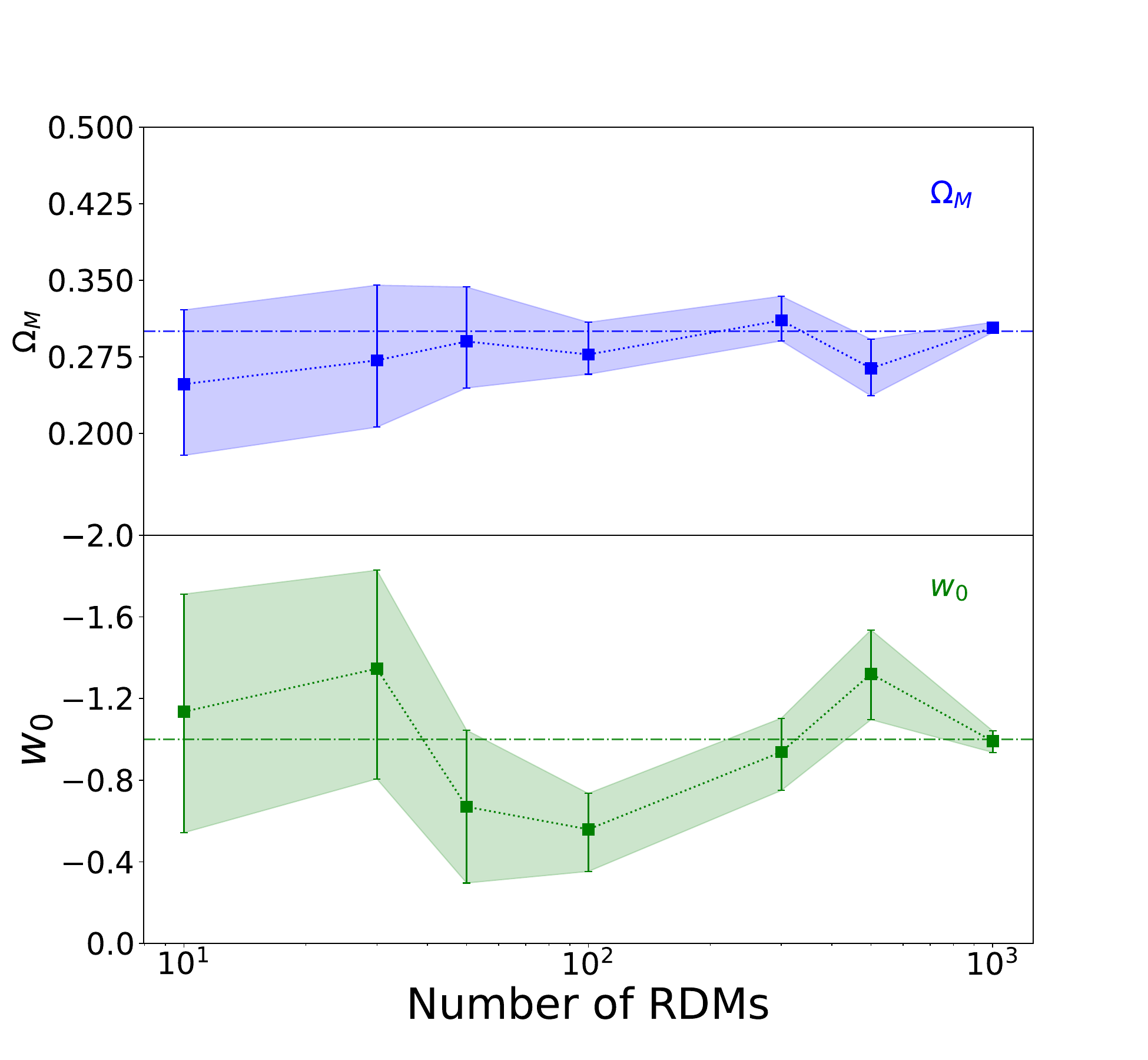}
    \caption{The graph depicts the evolution of errors in cosmological parameters for a flat $w$CDM universe model. In the upper panel, we observe the range of $\Omega_M$ for a varying number of clusters from 10 to 1000. In the lower panel, we see the changes in $w_0$ for cluster lenses fewer than 1000.}
    \label{fig: flat wCDM yerr}
\end{figure}

Beyond the standard $\Lambda$CDM model, we have applied the extended universe model where the equation of state parameter of dark energy differs from the $\Lambda$CDM model, in which the equation of state of dark energy is a constant $w_0$ \citep{PhysRevLett.90.091301Linder}. Recent CMB collaboration \citep{2020A&A...641A...6Planck} suggests the state parameter to be $w = -1.03 \pm 0.03$. The Hubble function in the $w$CDM model can be written as:
\begin{align}
    \begin{split}
        H(z) = H_0\left[ \Omega_M(1+z)^3 + \Omega_k(1+z)^2 + \Omega_\Lambda(1+z)^{3(1+w_0)} \right]^{1/2}
    \end{split}
    \label{eq: wCDM}
\end{align}
where $\Omega_M$, $\Omega_\Lambda$, and $w$ are the free parameters in MCMC sampling, with standard values of $\Omega_M = 0.3$, $\Omega_\Lambda = 0.7$, and $w = -1$. If we consider a flat universe, $\Omega_k = 0$ and $\Omega_M + \Omega_\Lambda = 1$, then there are only two independent variables: $\Omega_M$ and $w_0$. We sampled the distribution of the flat $w$CDM in Fig. \ref{fig: flat wCDM}, which gives $\Omega_M = 0.303\pm{0.005}$ and $w_0 = -0.99^{+0.055}_{-0.051}$. The errors of the flat $w$CDM are shown in Fig. \ref{fig: flat wCDM yerr}.

\label{sec: flat wCDM model}

\subsection{The $w\rm{CDM}$ model}
\label{sec:wCDM model}

\begin{figure}
    \centering
    \includegraphics[width=0.5\textwidth]{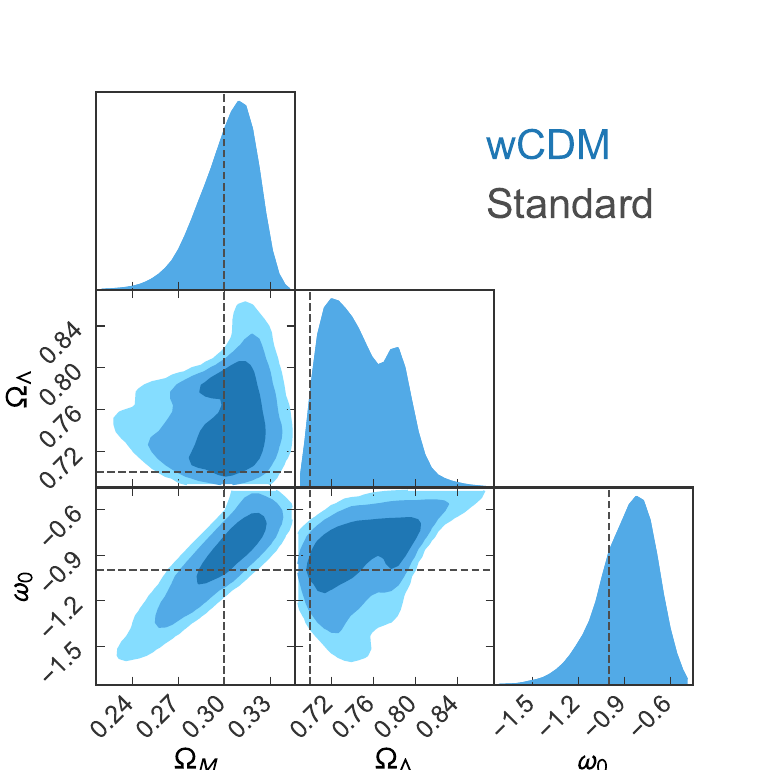}
    \caption{The MCMC result of the $w$CDM model indicates that the redshift difference limits the values of $\Omega_\Lambda$ to between 0.704 and 0.785, $\Omega_M$ to between 0.284 and 0.319, and $w_0$ to between -1.066 and -0.707.}
    \label{fig: wCDM}
\end{figure}
\begin{figure}
    \centering
    \includegraphics[width=0.5\textwidth]{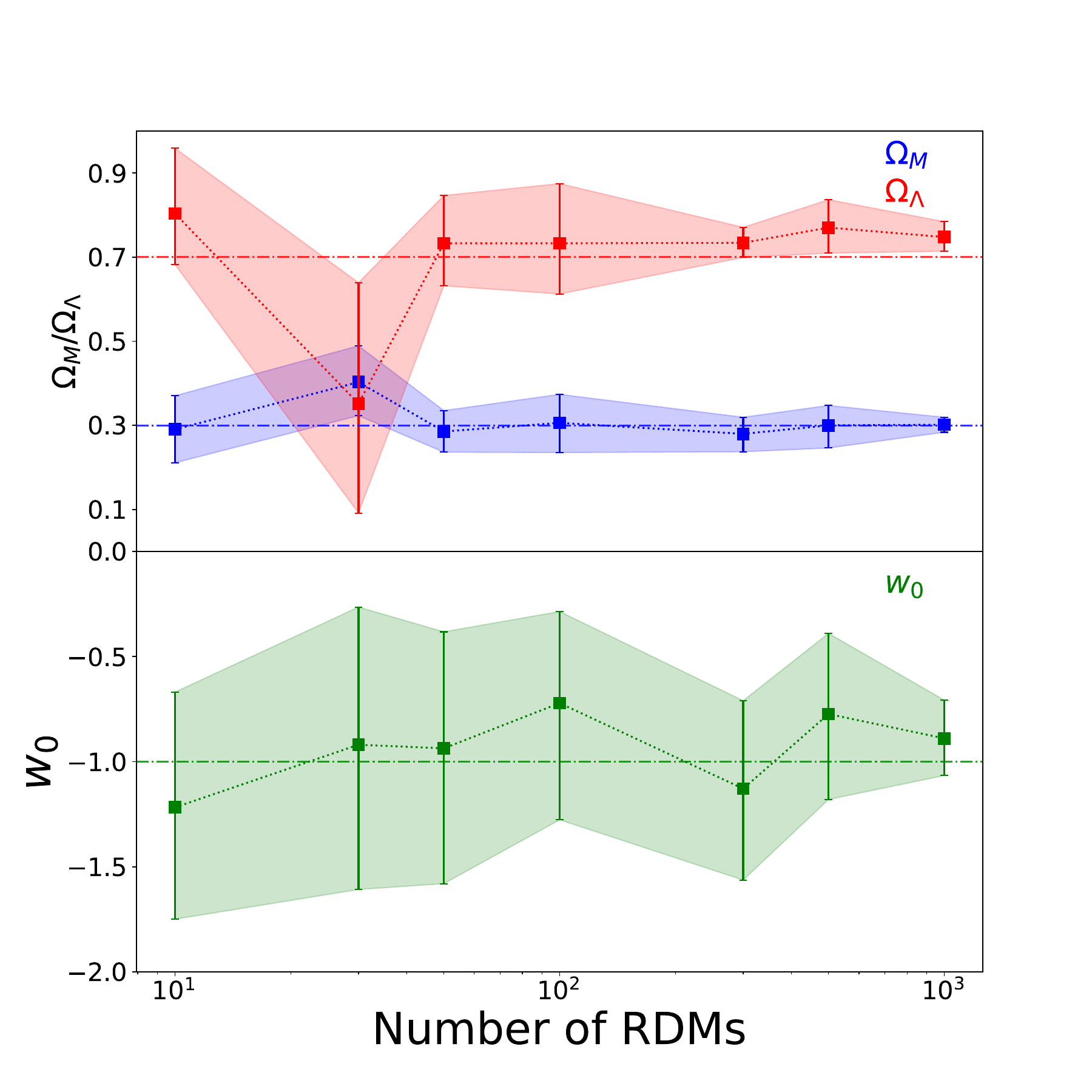}
    \caption{The figure constrains the tendency of cosmological parameter errors in a general $w$CDM universe model. In the upper panel, the red fillings are the change of the $\Omega_\Lambda$, and the blue colour refers to the $\Omega_M$. In the lower part, the green line refers to the error of $w_0$.}
    \label{fig: wCDM yerr}
\end{figure}

In the non-flat $w$CDM model, the MCMC result of sampling 1000 RDMs is presented in Fig. \ref{fig: wCDM}. The redshift difference places $68\%$ constraints on $\Omega_\Lambda$ in the range of $(0.704, 0.785)$, $\Omega_M$ in the range of $(0.284, 0.319)$, and $w_0$ in the range of $(-1.066, -0.707)$, which includes the standard model \citep{2020A&A...641A...6Planck}. In Fig. \ref{fig: wCDM yerr}, we show the error bars of these parameters for different numbers of clusters. This figure once again emphasizes the importance of measuring more than 1000 RDMs to accurately determine the cosmological parameters using the redshift difference. {Also, we notice that 100-RDMs case has the same constraint on $w_0$ as 1000-RDMs case. There is a degeneracy with $w\rm{CDM}$ when we involve more cosmological parameters. A possible explanation of this degeneracy is because we set a limited range of redshift of sources from $3$ to $6$. In this narrow redshift interval, as shown in Eqn. \ref{eq: wCDM}, a high $\Omega_\Lambda$ will have the similar results as a high equation of state $w_0$. This degeneracy could be broken if we include high-redshift sources, such as candidates reported by JWST with redshift $z > 10$ recently.\citep{2023MNRAS.518.4755Adams, 2020AAS...23520711L_high_redshift_JWST} }

\subsection{The Chevallier-Polarski-Linder (CPL) model}
\label{sec: CPL model}
\begin{figure}
    \centering
    \includegraphics[width=0.5\textwidth]{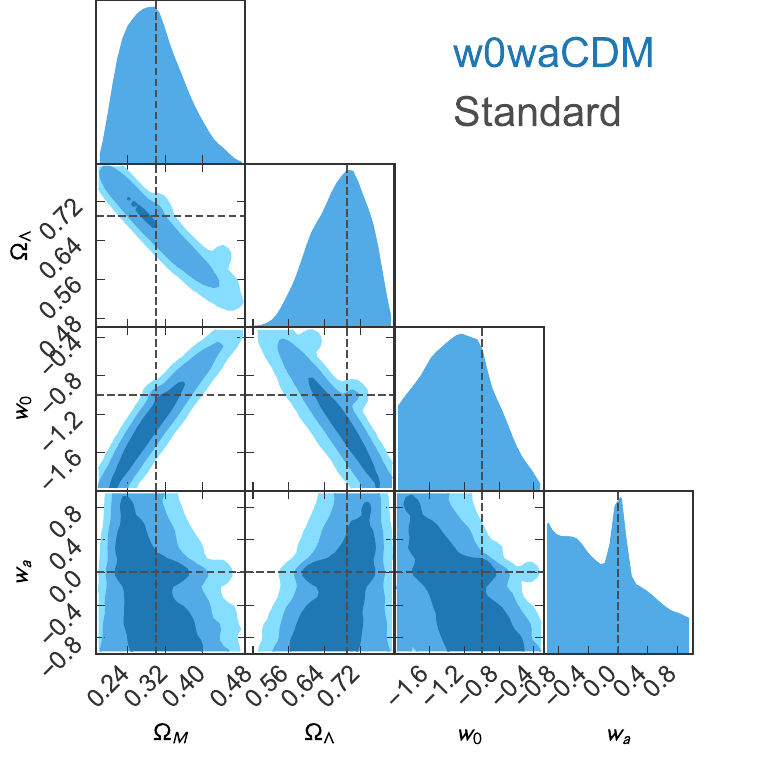}
    \caption{The MCMC sampling of the CPL cosmology shows that it provides strong constraints on the matter component and dark energy, while the constraints on the $w_0$ and $w_a$ parameters are weaker.}
    \label{fig: w0waCDM}
\end{figure}

\begin{figure}
    \centering
    \includegraphics[width=0.5\textwidth]{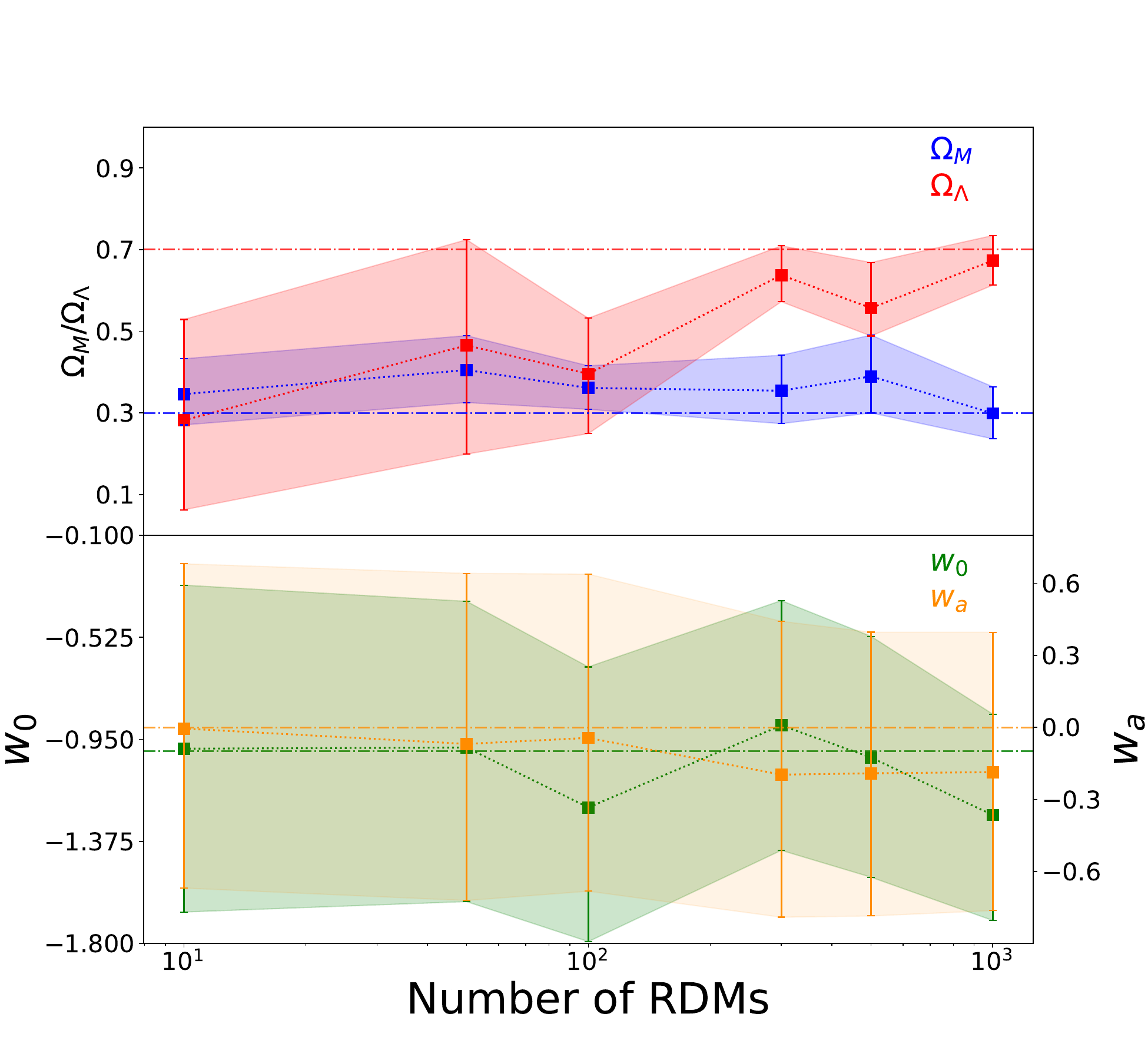}
    \caption{The contour plots in Fig. \ref{fig: w0waCDM} depict the $68\%$ confidence level of the cosmological parameters in a $w_0 w_a$CDM universe. The upper panel shows the constraints on $\Omega_\Lambda$ (red contour) and $\Omega_M$ (blue contour), while the lower panel shows the errors on $w_0$ and $w_a$.}
    \label{fig: w0waCDM yerr}
\end{figure}

Another parameterized model of the equation of state is the CPL model where the parameter of dark energy is supposed to be \citep{PhysRevLett.90.091301Linder}
\begin{align}
    \begin{split}
        w = w_0 + w_a \frac{z}{1+z}.
    \end{split}
\end{align}
This model performs higher accuracy at high redshift than many other scalar equations of field. The Hubble function of the CPL model is written as
\begin{align}
    \begin{split}
        H(z) = H_0& \left[ \Omega_M(1+z)^3 + \Omega_k(1+z)^2 \right. \\
         &+ \left. \Omega_\Lambda(1+z)^{3(1+w_0+w_a)}\rm{exp}\left(\frac{-3w_a z}{1+z}\right) \right]^{1/2}.
    \end{split}
\end{align}
Fig. \ref{fig: w0waCDM} presents the MCMC sampling results of the CPL cosmology, which shows that this model is well-constrained on the matter component and dark energy, but has weaker constraints on the $w_0$ and $w_a$ parameters. In Fig. \ref{fig: w0waCDM yerr}, we display the error bars of the parameters for different numbers of clusters. {The similar degeneracy problem happens on $w_0$ and $w_a$ in the CPL model as in the $w$CDM model. For instance, the redshift difference does not impose excessive restrictions on the $w_a$ parameter.}

\section{SUMMARY AND PROSPECTS}
\label{sec: summary}
In this paper, we present a novel method for determining the history of the universe using the redshift differences arising from slight variations in multiple images of a gravitational lens system. While such small residuals were previously negligible in observations, the next-generation telescope, as discussed in \cite{2021Msngr.182...27M_HIRES}, presents an opportunity to distinguish these redshift differences. The redshift difference method has several advantages: (1) it provides a direct measurement of cosmological history without any other assumptions, (2) it can cover a wide range of redshifts, {(3) compared to the long-term measurement of redshift drift \citep{1961ApJ...133..355S_Sandage, Loeb_1998}, observing the redshift difference does not require a long time baseline for the observer frame. On the other hand, there are some disadvantages for the redshift difference: (1) For redshift drift, a single quasar probes several independent absorption lines simultaneously across a wide range of redshift, but it is not easy to approach for redshift difference; (2) The choice of lens model is a crucial factor that can introduce systematic errors in measurements.}\par

Initially, we provide the general equations of the redshift difference in Eqn. \ref{eq: redshift difference}. The redshift difference is proportional to the Hubble function, the angular diameter distance of the lens and source, but inversely proportional to the angular diameter distance from the lens to the source. This value depends on the lens model used. To determine the value of redshift difference, we apply the non-singular ellipsoidal (NIE) model \citep{1989MNRAS.238...43K_Christopher,1994A&A...284..285Kormann} in the form of Eqn. \ref{eq: surface density}, which is one of the very general symmetric lenses. In this work, the NIE lens models are constructed using \texttt{lenstronomy} \citep{2018PDU....22..189Blenstronomy}. Additionally, we discuss the influences of peculiar velocities. As we are only interested in the difference of redshift between images, only the second-order influences for the source and observer. The lens' peculiar velocity will alter the critical curves of the lensing. \cite{1983Natur.302..315Birkinshaw} proposed a formula for redshift shift in a moving lens that is aimed to solve the redshift difference due to the lens's peculiar velocity.\par

After obtaining the equation for redshift difference, we can use \texttt{emcee} \citep{Foreman_Mackey_2013emcee} to sample cosmological parameters, such as the matter component $\Omega_M$, dark energy component $\Omega_\Lambda$. In the second part of this paper, we sample cosmological parameters in several different models, including the flat $\rm{CDM}$ model, the $\Lambda\rm{CDM}$ model, the $w\rm{CDM}$ model, and the CPL model. We then discuss the distribution of cosmological parameters in each universe model. The redshift difference has strong constraints on $\Omega_M$ and $\Omega_\Lambda$. Specifically, in a $\Lambda\rm{CDM}$ model, the matter and dark energy components fit the standard model. Additionally, in the $w\rm{CDM}$ universe, the $68 \%$ confidence estimation of $\Omega_M$, $\Omega_\Lambda$, and $w_0$ are discussed in Sec. \ref{sec:wCDM model}. The sampling result of the CPL model is shown in Fig. \ref{fig: w0waCDM}. Besides the restrictions on cosmological parameters, we also estimate the number of targets we need to measure. As shown in Fig. \ref{fig: LCDM cluster number}, the errors of $\Omega_M$ and $\Omega_\Lambda$ become smaller if we measure more than $100$ RDMs.\par

Additionally, we plan to investigate the impact of systematic errors on our measurements . Systematic errors can come from various sources, such as uncertainties in the lens model or observational errors. We also aim to explore the possibility of combining redshift difference measurements with other cosmological probes, such as the cosmic microwave background and supernova data, to obtain tighter constraints on cosmological parameters. Finally, we will analyze the feasibility of using the redshift difference method to distinguish between different dark energy models, such as scalar field models or modified gravity models. These investigations will help us to better understand the universe's history and provide more accurate constraints on cosmological parameters.

\section*{ACKNOWLEDGEMENTS}
This research made use of \texttt{Astropy},\footnote{http://www.astropy.org} a community-developed core Python package for Astronomy \citep{2013A&A...558A..33AstropyI,2018AJ....156..123AstropyII}, \texttt{emcee} \citep{Foreman_Mackey_2013emcee} \footnote{https://github.com/dfm/emcee}, \texttt{lenstronomy} \footnote{ https://github.com/sibirrer/lenstronomy} \citep{2018PDU....22..189Blenstronomy}, \texttt{hmf} \citep{2013A&C.....3...23Murray} \footnote{https://github.com/halomod/hmf/blob/master/docs/index.rst} and \texttt{bagpipes} \footnote{https://bagpipes.readthedocs.io/en/latest/}. The corner plots are plotted by \texttt{corner} \citep{corner} and \texttt{pygtc} \citep{Bocquet2016pygtc}.\par

\section*{DATA AVAILABILITY}
The data underlying this article were generated from \texttt{lenstronomy} \citep{2018PDU....22..189Blenstronomy}. The derived data generated in this research will be shared upon reasonable request to the corresponding author.



\bibliographystyle{mnras}
\bibliography{Redshift_Difference_Cosmology} 




\appendix


\bsp	
\label{lastpage}
\end{document}